\documentclass[cpp,a4paper,fleqn%
]{w-art}
\usepackage{times,cite,w-thm}
\theoremstyle{plain}

\theoremstyle{definition}

\usepackage[]{graphicx}
\begin{document}
\DOIsuffix{theDOIsuffix}
\Volume{46}
\Month{01}
\Year{2007}
\pagespan{1}{}
\Receiveddate{September 23, 2013}



\title[Edge particle transport]{Guiding center orbit following calculation of edge particle
and heat transport in stochastic magnetic field}


\author{C.~C.~Chang, 
Y.~Nishimura
\footnote{Corresponding author\quad E-mail:~\textsf{nishimura@pssc.ncku.edu.tw}}, 
{\rm and} C.~Z.~Cheng
}
\address{Institute of Space and Plasma Sciences, National Cheng Kung University, Tainan 70101, Taiwan}

\begin{abstract}
Particle and heat transport in tokamak edge is investigated by 
guiding center orbit following calculation.
The guiding center equation is solved for both electrons 
and ions in the presence of magnetic perturbation. 
It is suggested that the remnants of the magnetic islands play
a role in characterizing the radial transport.
The transport coefficient is estimated which also demonstrate
local structure in the vicinity of magnetic islands.
\end{abstract}
\maketitle                   





\section{Introduction}

\indent
Particle and heat transport in tokamak edge plays crucial roles in sustaining the plasma discharge. 
Externally imposed magnetic perturbation can induce magnetic stochasticity 
which can then prevent disruptive phenomena.\cite{eva06}  
While magnetic field lines exhibit magnetic stochasticity in the presence of incommensurate magnetic perturbations,
corresponding particle motions are also influenced by mirror force and perturbed electric fields. 
 
The externally imposed Resonant Magnetic Perturbations (RMP)\cite{eva06} degrades ion and electron particle transport 
while the electron heat transport does not change significantly 
(the electron mass transport component does not appear in the heat transport to
the extent one expects from the density gradient\cite{moy13}) and ion temperature rather rises.
The latter feature is inconsistent with the Fick's law and Fourier's law in one dimensional diffusion model.
These puzzling features are further confirmed in recent DIII-D experiments.\cite{nf11,nf12}
A few possible ingredients are considered in this work to account for the inconsistency:
(1) magnetic field lines are not completely stochastic\cite{rec78} 
and the remnants of magnetic islands play a role, (2) trapped particles effects 
due to the equilibrium magnetic field inhomogeneity, (3) radial electric field and convective transport.\cite{wes97} 
To investigate these latter effects, we employ a guiding center particle model.

Being different from self-generated tearing instabilities,\cite{fur63}
in the RMP experiments, magnetic perturbations are given
by the external magnetic I-coils.
From a computational point of view, the stochastic components of the magnetic field 
can be given time independent together with the equilibrium magnetic field.
In this work, we would like to develop a numerical tool
to investigate the tokamak edge transport.
The guiding center equation is solved in the presence of magnetic perturbation. 
We would like to see the direct response of particles subject to the RMP.
We then discuss density and temperature evolution by sampling the particles from
the orbit calculation.


\section{Computational model}
\label{s2}
The guiding center equation in a pseudo-toroidal coordinate\cite{miy86} is given. 
Normalizing the length by the minor radius $a$, 
time by the inverse of the cyclotron frequency of electrons $\Omega_{ce}$ (or ions, $\Omega_{ci}$),
and the magnetic field strength at the magnetic axis $B_0$,
the guiding center equation components in the normalized units are given by
\begin{equation}
dr/dt =  - \varepsilon \mu \sin{(\theta)} 
- v_{\|} \sum_{m} b_{rm} \sin{ \left(  m \theta - n \zeta \right) },
\label{gc1a}
\end{equation}
\begin{equation}
d \theta/dt = \varepsilon v_\| /q(r)  - ( \varepsilon \mu/ r ) \cos{(\theta)} ,
\label{gc1b}
\end{equation}
\begin{equation}
d \zeta/ dt  = \varepsilon v_\|,
\label{gc1c}
\end{equation}
\begin{equation}
d v_{\|}/ dt = - [ \varepsilon^2 \mu r / q(r) ] \sin{(\theta)} .
\label{gc1d}
\end{equation}
Here the normalized radial coordinate, the poloidal angle, and the toroidal angle 
are given by $r, \theta$, and $\zeta$ respectively. 
The parallel velocity is given by $v_{\|}$ and $\mu$ is the magnetic moment of the electrons and ions.
In Eqs.(1)-(4), the normalized magnetic field strength is given by
$B(r, \theta) = 1 - \varepsilon r \cos{\theta }$ where the inverse aspect ratio is given by $\varepsilon $.
The safety factor is given by $q(r)=1+3 r^3$. 
As in the RMP experiments\cite{eva06}, we have taken a single toroidal mode number $n=3$ and
poloidal mode numbers $9 \le m \le 12$ in the computation.
Each mode resonates at the mode rational surfaces $r_{m/n}$ where $q(r_{m/n}) = m/n$.
To time advance Eqs.(1)-(4), a fourth order Runge-Kutta-Gill method is employed.
The magnetic island width in the absence of the drift terms and the mirror force is given by 
$W_{m} = 2 \sqrt{b_{rm}q(r_{m/n})^2} / \sqrt{dq/dr(r_{m/n}) \varepsilon m  }$.
Here, $W_{m}$ is independent of parallel velocity since (passing) particles freely stream 
along stochastic magnetic field lines.


\section{Computational results}
\label{s3}

We investigate the evolution of the density and the temperature for the electrons and ions.
Parameters used in the calculation are similar to the DIII-D tokamak,\cite{eva06,nf11,nf12}
major radius $ {R}_{0} =  3.0 m$, minor radius $ a =  1.0 m$,
toroidal magnetic field strength $ {B}_{0} = 2 T$.
We have taken the temperature at the pedestal top to be $ {T}_{i} = {T}_{e} = 1  keV$.
The RMP amplitudes are given by 
$B_r/B_0= 1.0 \times 10^{-3}$ ($B_r$ is the radial component of magnetic perturbations)
for all the $9 \le m \le 12$ ($n=3$) modes
which are about three times larger than the RMP experiments\cite{eva06}
(note that our safety factor profile is not as sharp as Ref.\cite{eva06}
and we took larger $B_r$ values to have islands overlapped).
Poincar\'{e} mapping of magnetic field lines are shown in Fig.~1(a) and Fig.~2(a).
Figure 1(a) incorporates all the $9 \le m \le 12$ modes
while Fig.~2(a) retains only the $m= 9$ mode to demonstrate the formation of
a single magnetic island chain.
Note that in the RMP experiments, the I-coil fields\cite{eva06,nf11,nf12} 
are externally imposed ones and differ from self-generated tearing modes.\cite{fur63}
In Fig.1(a), one can observe significant amount of the island remnants.
As a reference, the measured magnetic diffusion levels are $D_{M} = 1.48 \times 10^{-6} m$ at $r/a=0.80$,
$D_{M} = 3.06 \times 10^{-6} m$ at $r/a=0.85$, and
$D_{M} = 3.36 \times 10^{-6} m$ at $r/a=0.90$
(Ref.\cite{eva06} provides quasi-linear diffusion of $D^{ql} = 3.5 \times 10^{-6} m$).

Initial condition of the particles are given
to satisfy the equilibrium density profile of $n (r) = n_0 [1 - (r/a)^4]$
and the temperature profile of $T (r) = T_0 [1 - (r/a)^4]$ for both the ions and electrons.
Here $n_0$ and $T_0$ are the density and temperature at the magnetic axis.
On each magnetic surface, the test particles are uniformly set in the poloidal angle on the $\zeta=0$ plane.
The velocity distribution is given by a bi-Maxwellian using a random number generator.\cite{nis11}
A total of 1,582,700 particles are used in the orbit following calculation.

When the particles reach $r/a=1.0$ we remove them from the computation.
We presume they are lost by reaching the divertor and 
then to the open-field-line scrape-off-layer regions.
The divertor geometry is not included in this work, however.
We only deal with closed magnetic field lines in this paper.
Toward the core region, one finds Kolmogorov-Arnold-Moser (KAM) surface [near $r/a=0.75$ in Fig.1(a)]
which behaves as an inner reflection boundary for the particles. 
In the orbit following calculation, we set initial particles in the radial domain $0 \le r/a \le 1$, however.
To estimate the density and the temperature evolution, we evolve the guiding center
equation and after finite time assign particles into radially uniform bins. 
When the particles are in between mesh points,
the particle density is assigned onto neighboring mesh points by a linear interpolation. 
Five hundred mesh points are taken within $0 \le r/a \le 1$. 
Unless there is a net radial transport (Coulomb collisional or stochasticity induced transport), 
the number of the particles conserves.
For the temperature estimation, we take the second order velocity moments of the each particle
and assign them into the radial bins.

Figure 1 (b) shows the electron density $n_e (r)$ evolution and
Fig.1 (c) shows the electron temperature $T_e (r)$ evolution which are obtained from 
the guiding center calculation (in the presence of the RMP imposed stochastic magnetic field).
The passing electrons are freely streaming along stochastic magnetic field lines.
The trapped particles experience banana motion but they do not close on itself due to the perturbation
(different from banana orbits in completely axisymmetric tokamaks).
Since the magnetic field lines are stochastic, we observe effective radial diffusion.
Figure 1 (b) and Fig.1 (c) show radial stepwise structure which reflects the remnants of magnetic islands.
Since we have no source term but have a sink at $r/a = 1.0$, 
both the $n_e (r)$ and the $T_e (r)$ profiles decay with time (corresponds to solving diffusion
equation with a homogeneous Neumann boundary condition at $r/a \simeq 0.75$ 
and a Dirichlet boundary condition at $r/a=1.0$).

To clarify the reflection of the magnetic island in the particle dynamics, a single $m=9$ mode 
is incorporated in Fig.2. We observe flattening in the vicinity of the magnetic island.\cite{cha90}
The flattening is due to the trapping of the particles which rotate inside the magnetic island
(due to the rotation,\cite{nic92} the inboard side and the outboard side of the islands have approximately
the same passing particle population).
Note that what we see in Fig.1 (b)(c) and Fig.2 (b)(c) 
is merely a radial projection of the helically twisted three dimensional structure. 
On the other hand, note that the temperature evolution is due
to the transport of particles themselves and is convective
(conductive heat transport\cite{cha98} is not incorporated in our model).

\begin{figure}
\includegraphics[width=43mm,height=43mm]{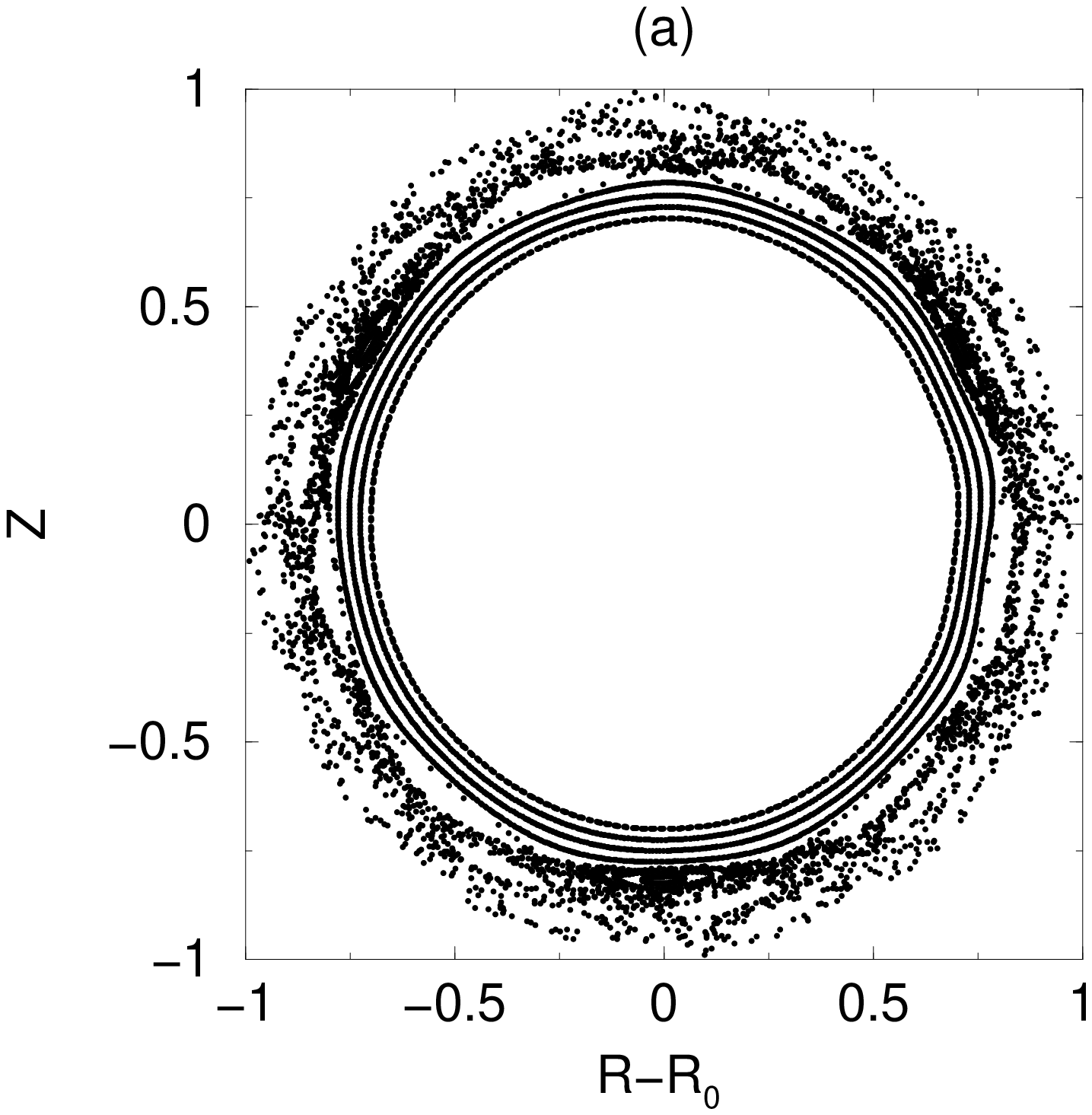}
\hfil
\includegraphics[width=43mm,height=43mm]{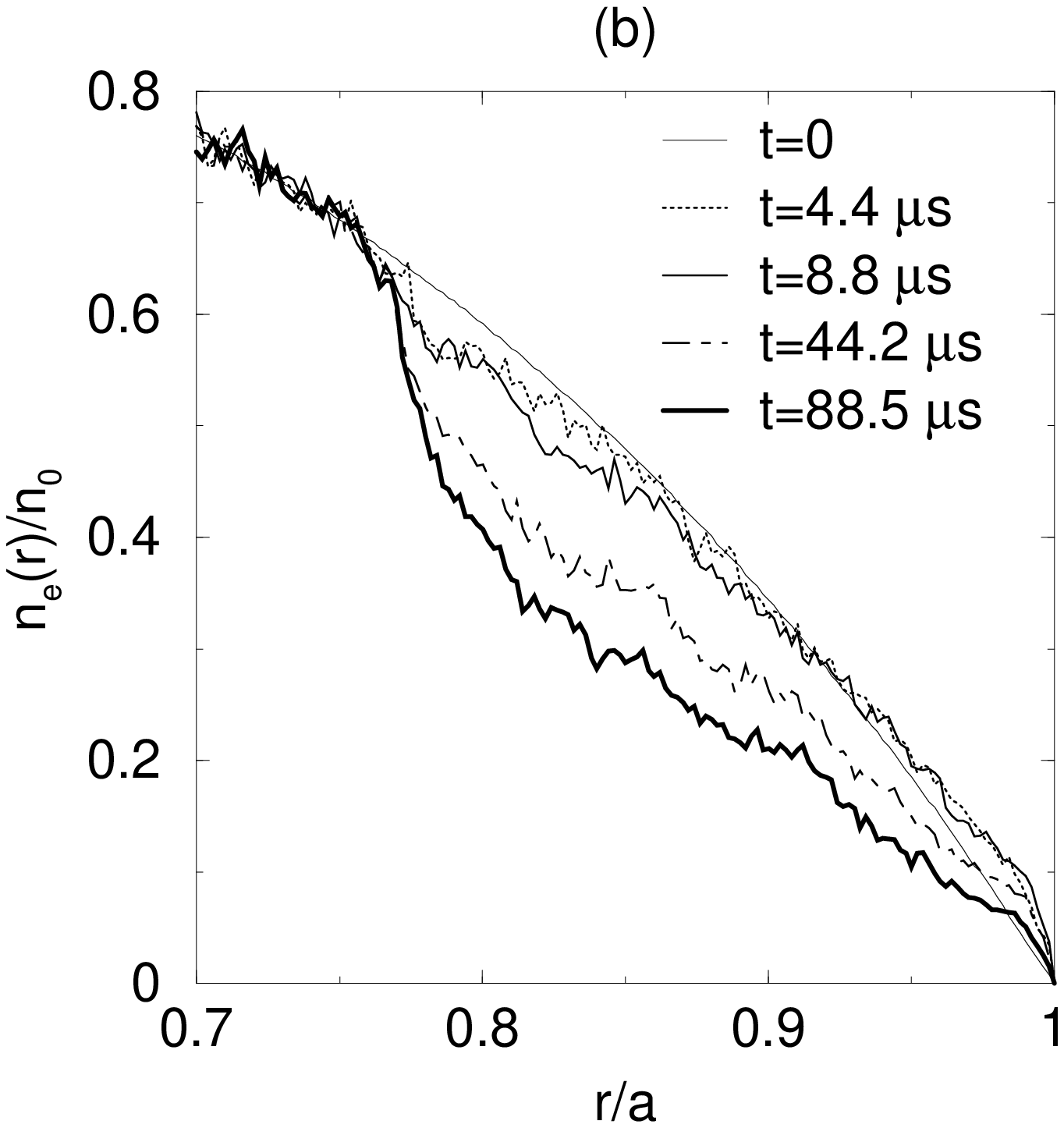}
\hfil
\includegraphics[width=43mm,height=43mm]{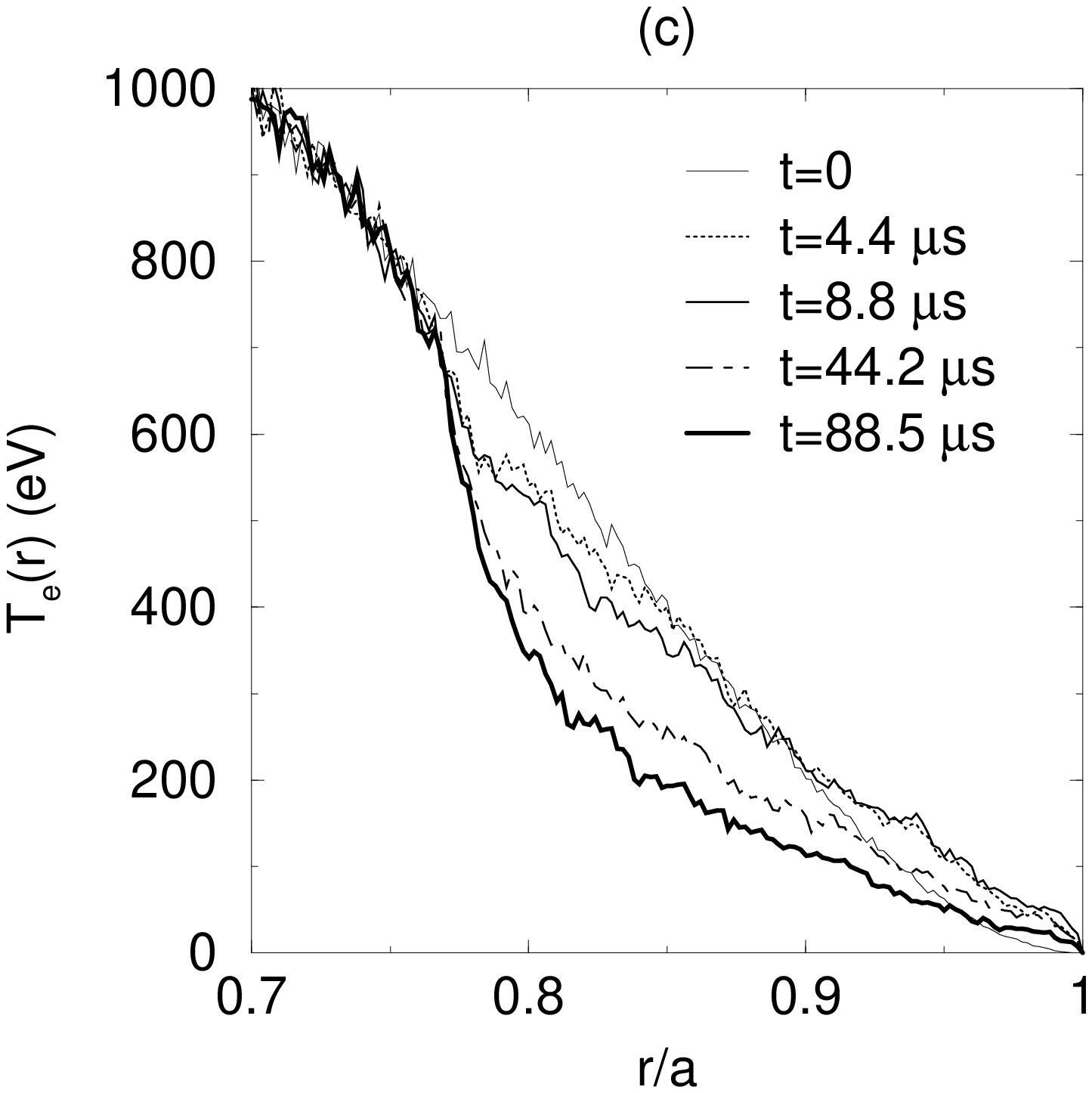}
\caption{(a) Poincar\'{e} plots of magnetic field lines in the presence of externally imposed
magnetic field at the $\zeta=0$ plane. Multiple poloidal modes ($9 \le m \le 12$)
employed. Corresponds to guiding center Poincar\'{e} 
plots in the absence of mirror force and gradient-B drift. 
The radial and vertical coordinates are given 
by $R - R_0$ and $Z$ respectively, where $R$ is the major radius and $R_0$ is
the major radius at the magnetic axis.
(b) Electron density evolution, and (c) electron temperature evolution versus minor radius.
Profiles are shown at $t=0 $, 
$t=4.4 (\mu s)$, $t=8.8 (\mu s)$, $t=44.2 (\mu s)$, and $t=88.5 (\mu s)$. 
Notice locally flattened structures due to the remnants of magnetic islands.}
\label{fig1}
\end{figure}
\begin{figure}
\includegraphics[width=43mm,height=43mm]{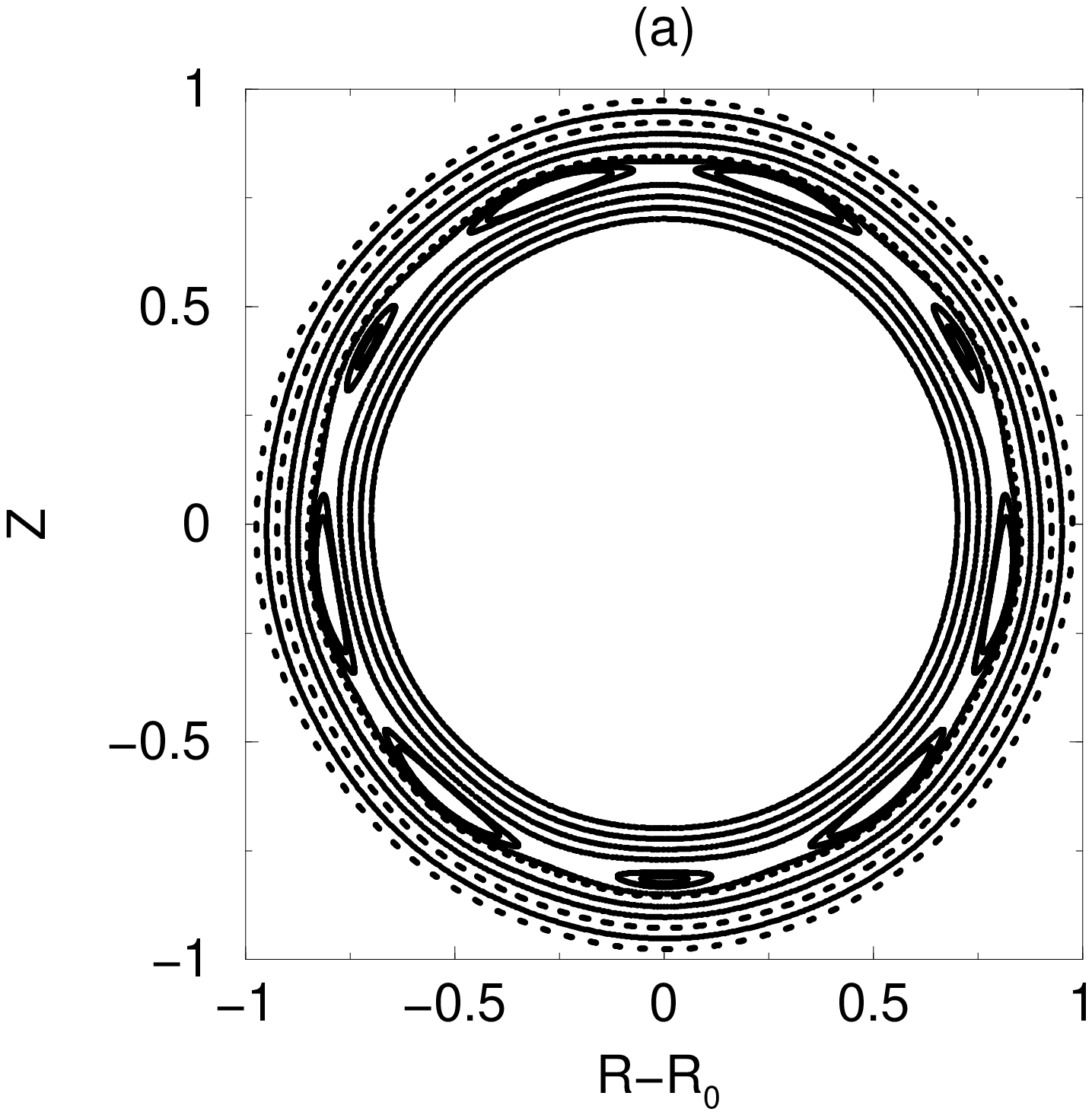}
\hfil
\includegraphics[width=43mm,height=43mm]{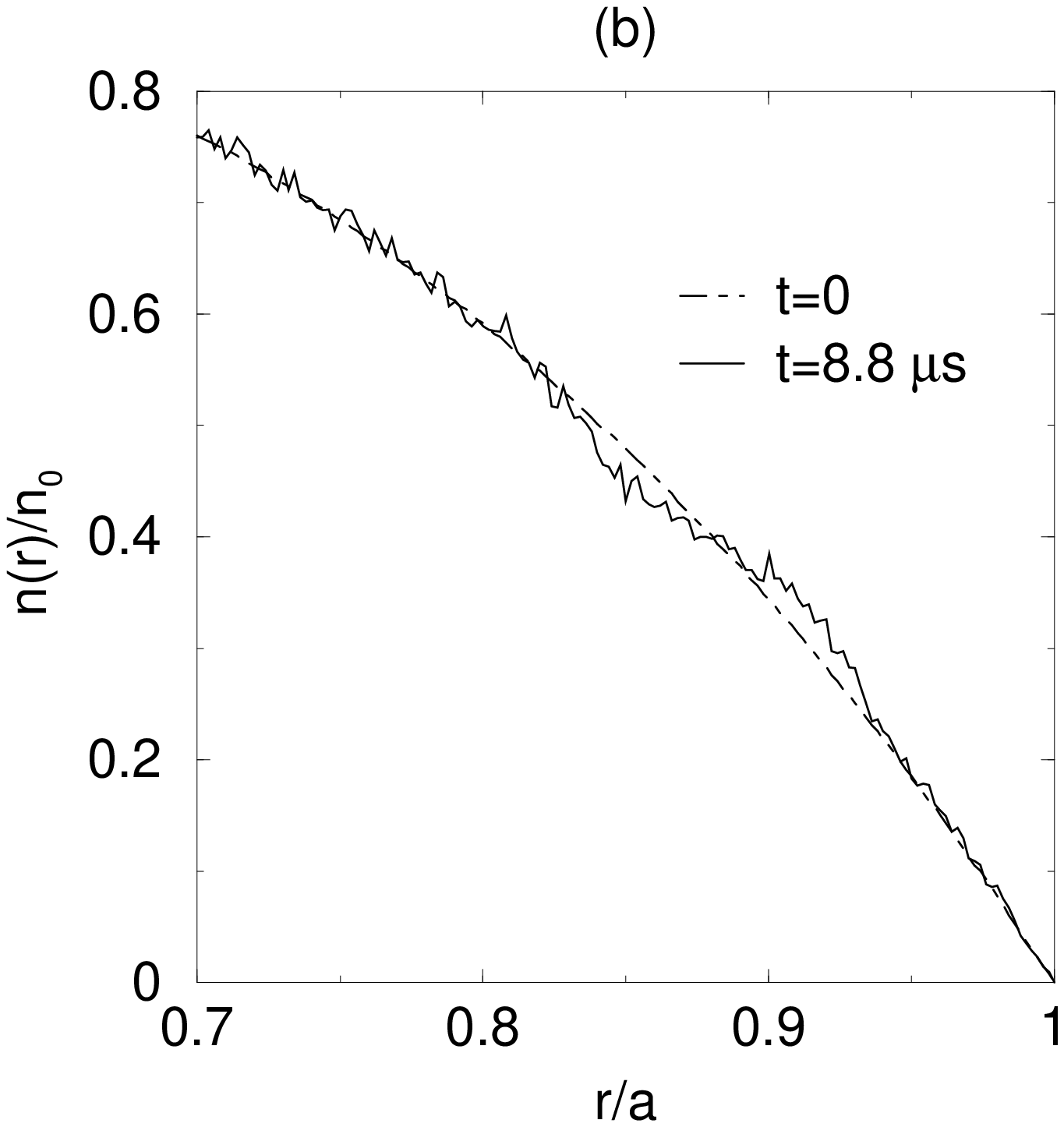}
\hfil
\includegraphics[width=43mm,height=43mm]{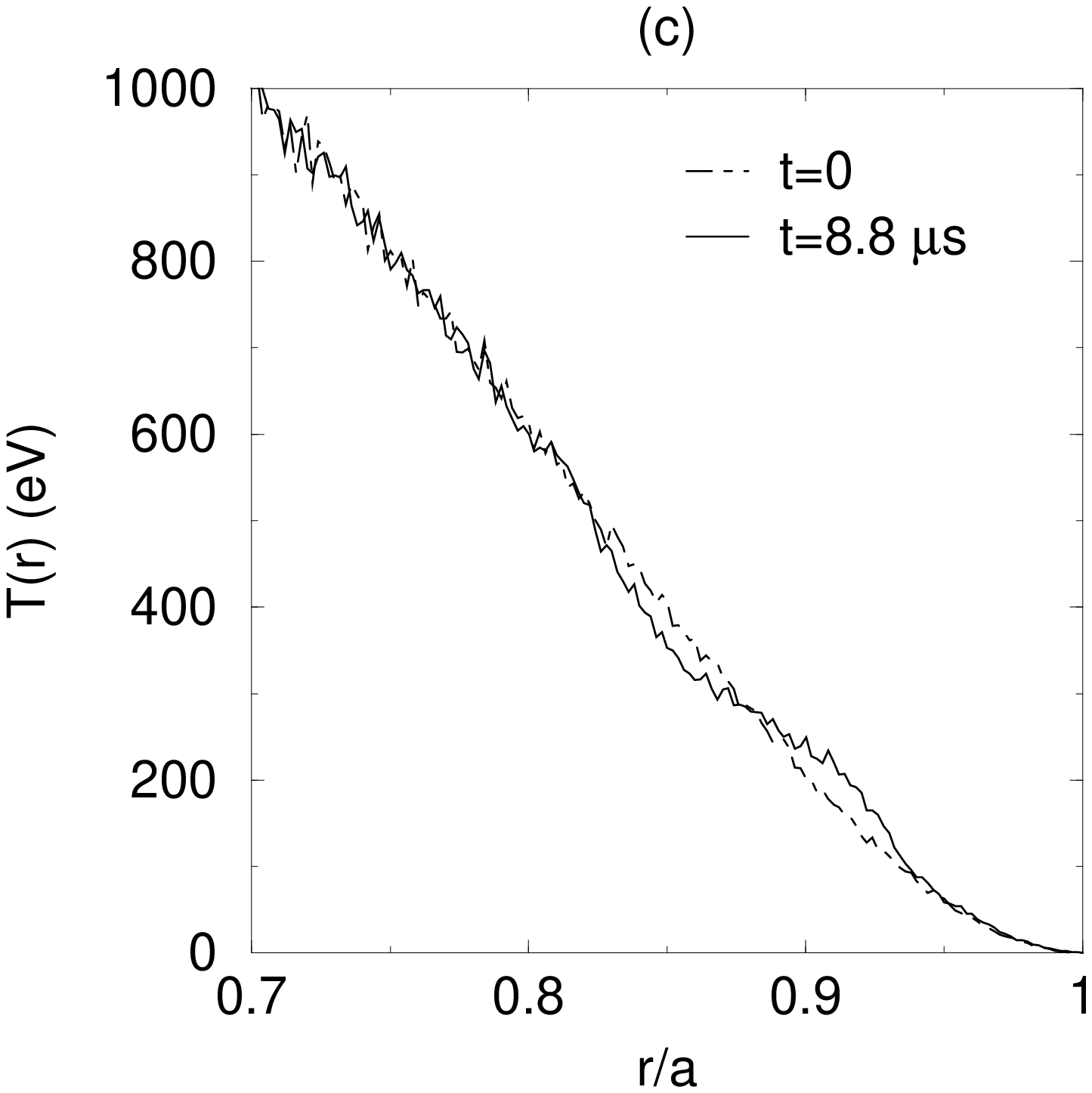}
\caption{(a) Poincar\'{e} plots of magnetic field lines in the presence of externally imposed
magnetic field at the $\zeta=0$ plane. A single poloidal mode ($m=9$) is employed. 
(b) Electron density evolution, and (c) electron temperature evolution.
Profiles are shown at $t=0 $ and $t=8.8 (\mu s)$. 
Observe the flattened structure in the vicinity of $0.8 \le r/a \le 0.85$.
The flattened structure is due to the trapping of the passing particles by
the $m=9$, $n=3$ magnetic island.}
\label{fig2}
\end{figure}
A series of computation in Fig.3 demonstrates evolution of ion density $n_i (r)$. 
Figure 4 is the evolution of the ion temperature $T_i (r)$, correspondingly.
We have employed deuterium for the ion species. 
Figure 3(a) employs the same geometrical and field parameters with Fig.1. 
Note that for $T_i = 1000 (keV)$ deuteriums, the banana width can be as large as a few percent
of the minor radius. The time scale and the finite banana width are the major differences
from the electron case.
As in the electron case, KAM surface near $r/a \simeq 0.75$ behaves as a reflecting boundary. 
Both the ion density $n_i (r)$ and the ion temperature $T_i (r)$ decays with time.
As for the electrons, a stepwise structure is found corresponding to the island remnants.

In Fig.3(b), we raise the amplitude of the magnetic perturbation.
The magnetic perturbation is given by $B_{r}/B_0 = 4.0 \times 10^{-3}$ 
(increased by a factor of four from Fig.3(a), thus each
magnetic island is twice larger).
Note that the KAM surface is slightly shifted inward ($r/a \simeq 0.7$) compared to Fig.3(a).
The stepwise structure disappears accompanied by enhanced particle stochasticity and much rapid profile evolution.
For Fig.3(a) and Fig.3(b), the inverse aspect ratio of $\varepsilon=0.33$ is taken.

In Fig.~3(c), we vary the inverse aspect ratio $\varepsilon$ and compare the time scale of the 
profile evolution with Fig.~3(a).
The long dashed curve is for $\varepsilon=0.5$ and the thick solid curve is 
for $\varepsilon=0.33$.
Both the curves are at $t=5.36 ms$ [see the thick solid curve in Fig.3(a) at $t=5.36 ms$].
The magnetic perturbation of $B_{r}/B_0 = 1.23 \times 10^{-3}$ is taken
(to have the same magnetic island widths which depends not only on $B_r$ but also on $\varepsilon$).
Otherwise we take the same parameter as in Fig.3 (a).
By the increase of the trapped particles' fraction, we find the transport level is reduced.
By diagnosing single particle orbits, we found that the radial excursion
of trapped particles is reduced.
We conjecture the reduction of the transport level
is due to the conservation of second adiabatic invariant for the trapped particles.
\begin{figure}
\includegraphics[width=43mm,height=43mm]{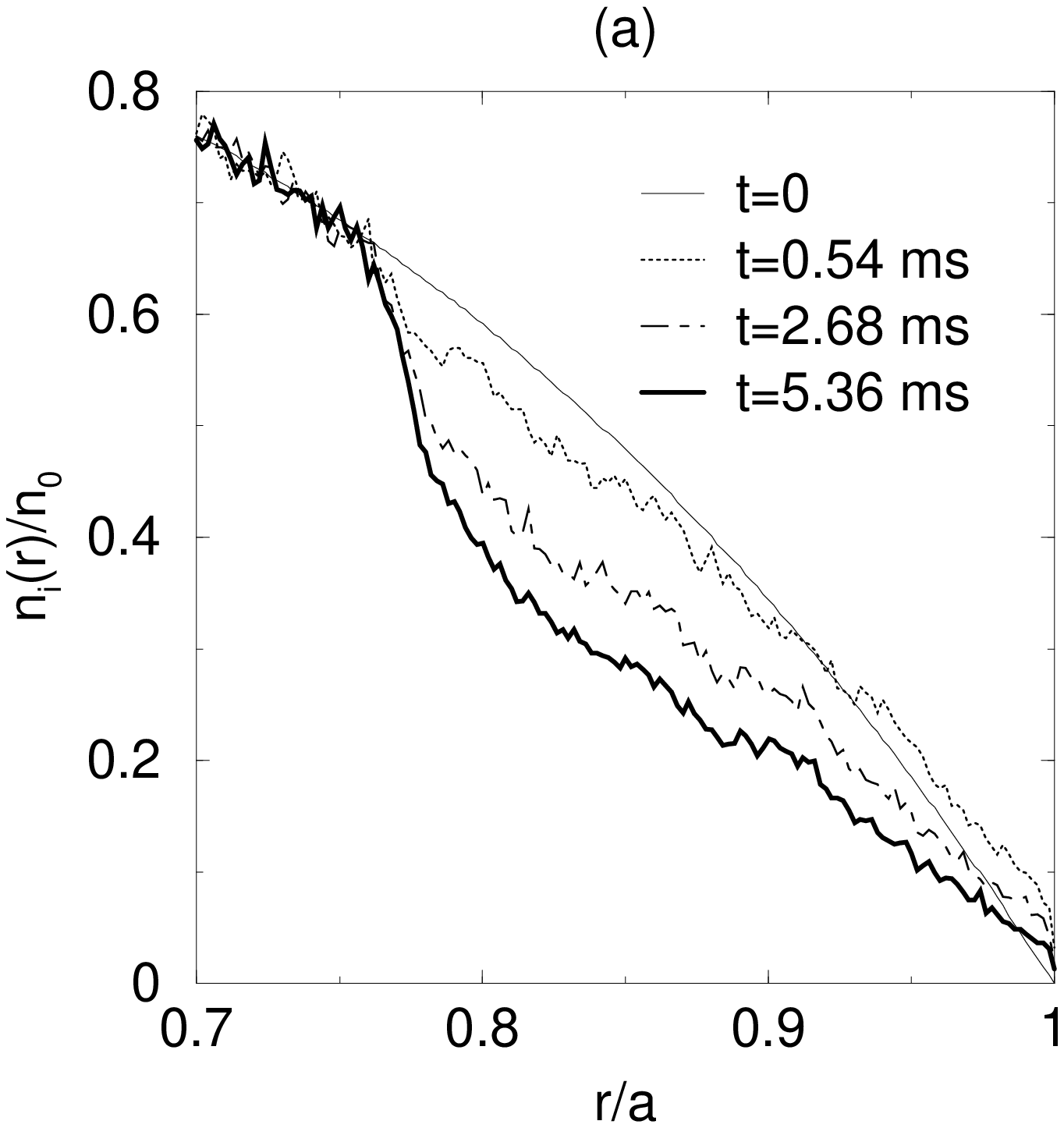}
\hfil
\includegraphics[width=43mm,height=43mm]{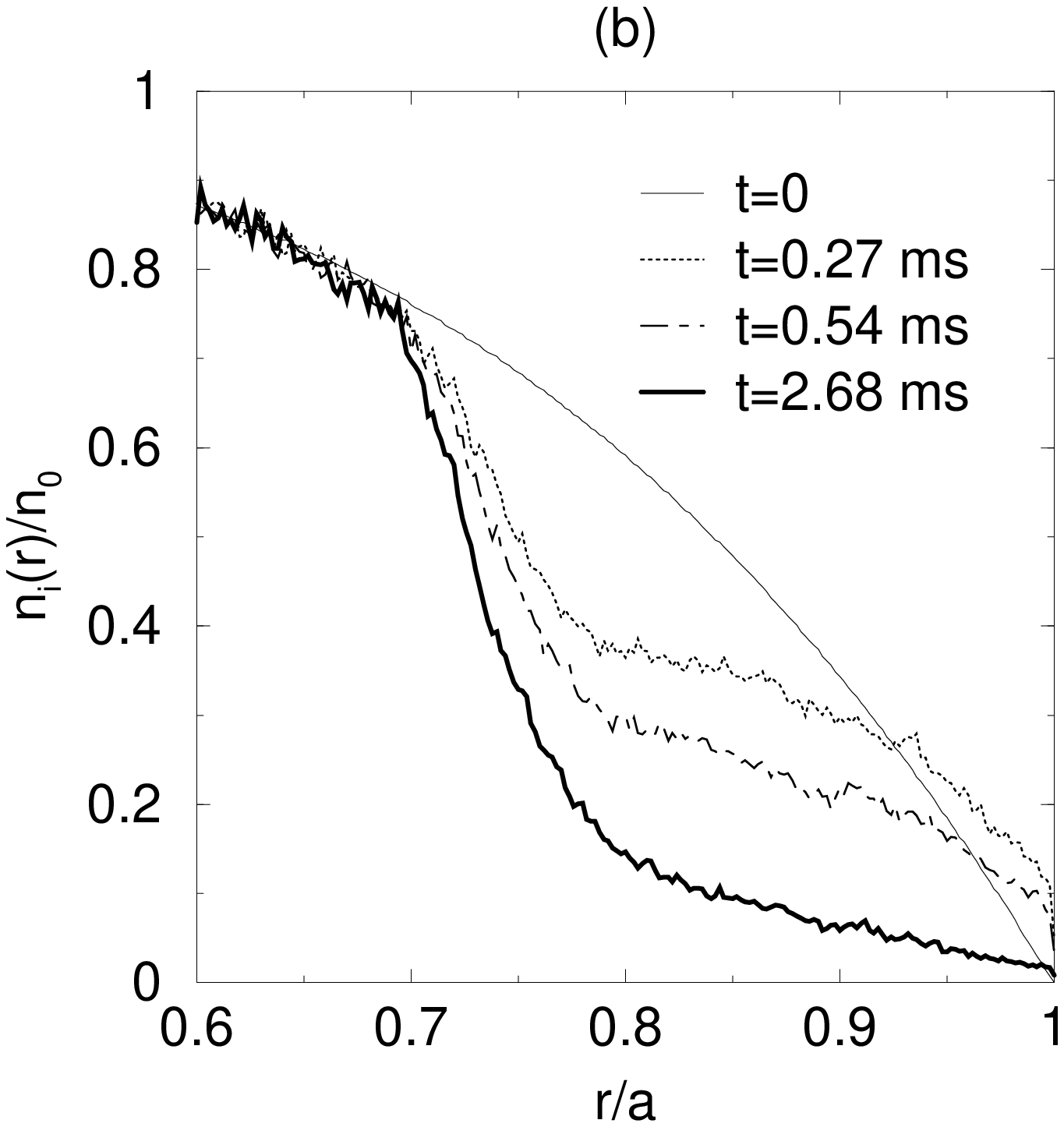}
\hfil
\includegraphics[width=43mm,height=43mm]{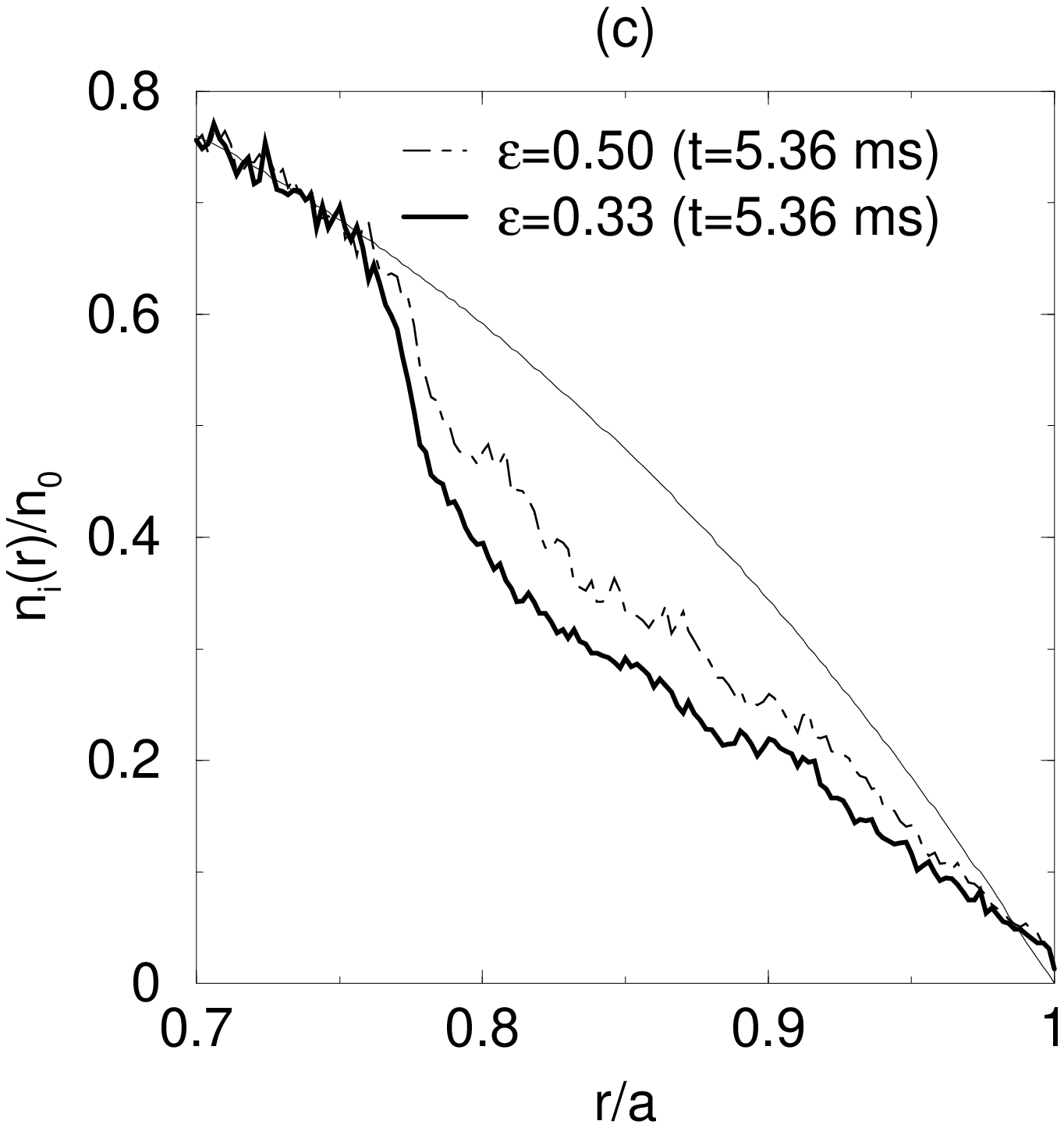}
\caption{Ion density evolution in the presence of $9 \le m \le 12$ ($n=3$) modes.
(a) The same parameters as in Fig.1 taken with profiles shown at $t=0 $, 
$t=0.54 (ms)$,  $t=2.68 (ms)$, and $t=5.36 (ms)$,
(b) the magnetic perturbation is given by $B_{r}/B_0 = 4.0 \times 10^{-3}$
with profiles shown at $t=0 $, 
$t=0.27 (ms)$,  $t=0.54 (ms)$, and $t=2.68 (ms)$,
and (c) the inverse aspect ratio of $\varepsilon=0.5$ is taken and compared with the $\varepsilon=0.33$ case
at $t=0.536 (ms)$. 
}
\label{fig3}
\end{figure}
\begin{figure}
\includegraphics[width=43mm,height=43mm]{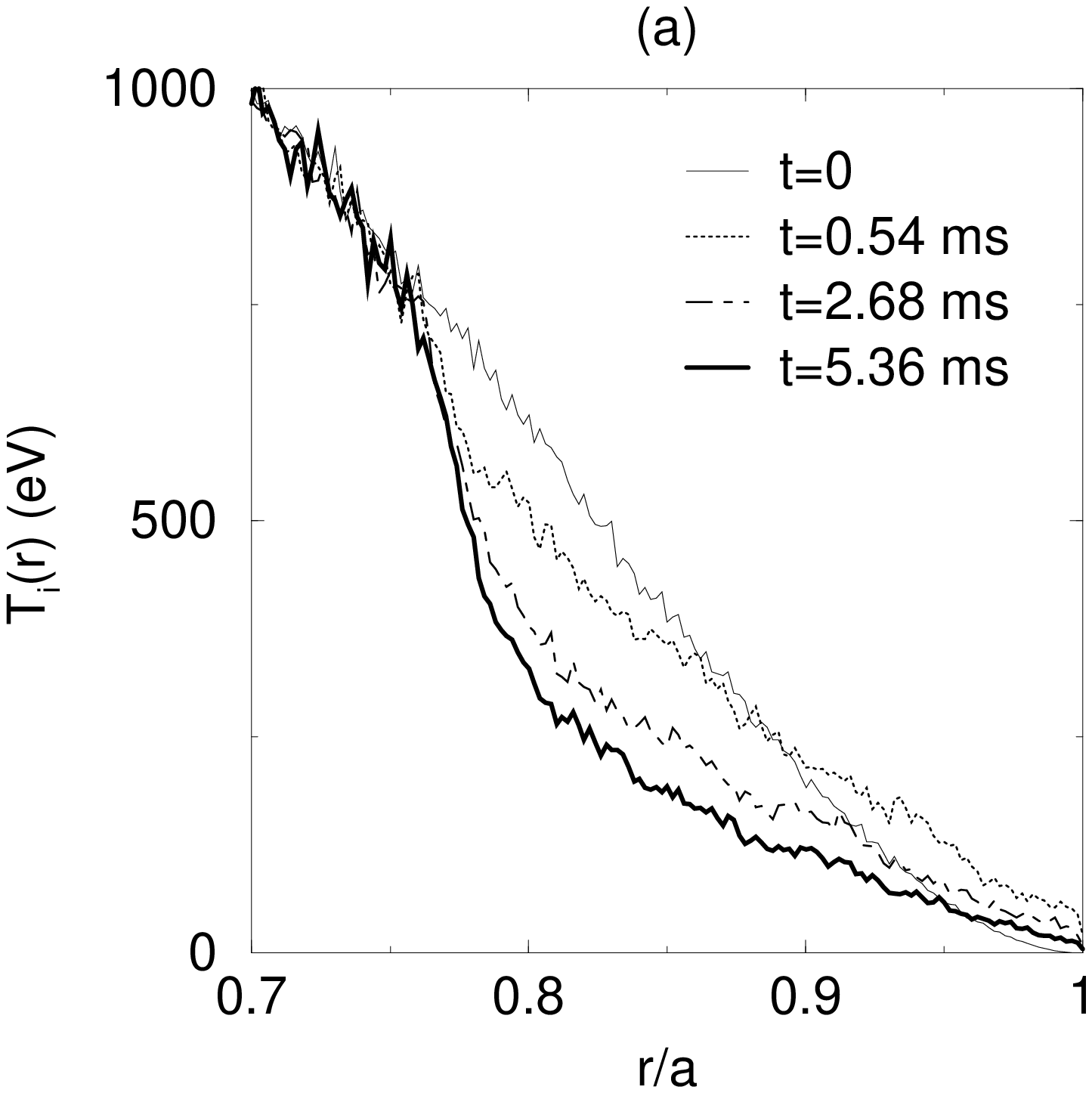}
\hfil
\includegraphics[width=43mm,height=43mm]{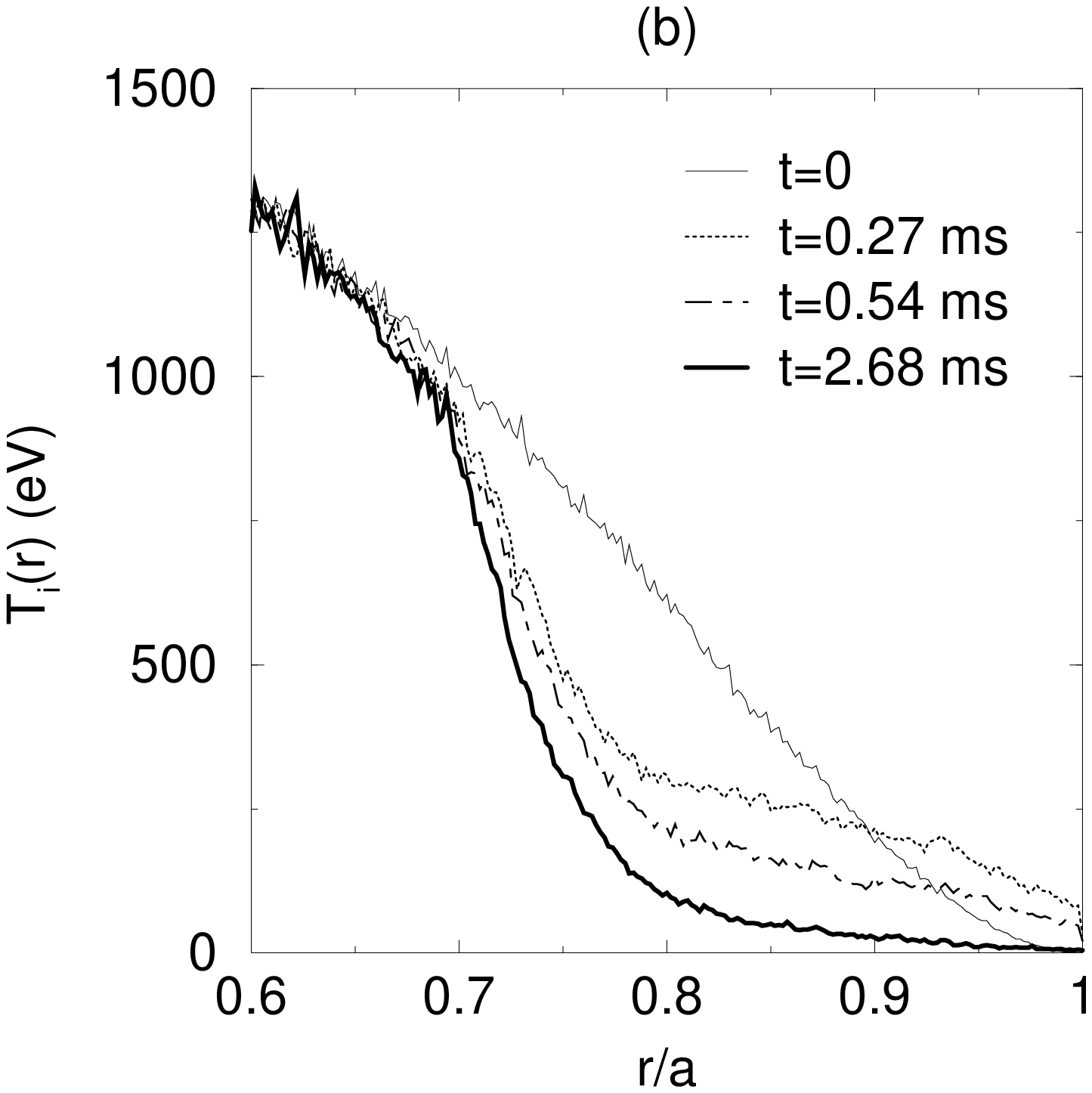}
\hfil
\includegraphics[width=43mm,height=43mm]{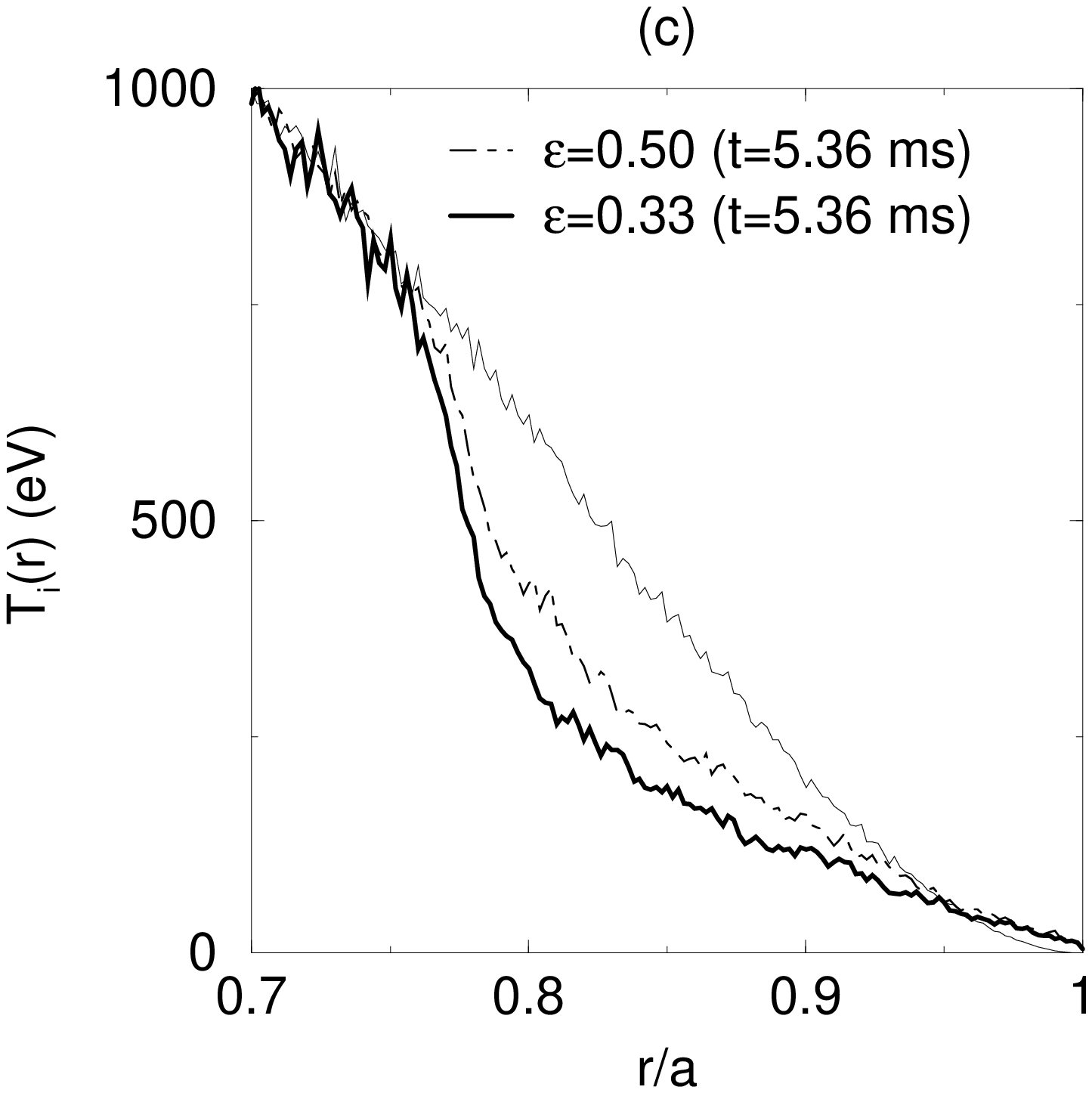}
\caption{Ion temperature evolution in the presence of $9 \le m \le 12$ ($n=3$) modes.
Corresponds to (a), (b), and (c) of Fig.3. }
\label{fig4}
\end{figure}

Local electron and ion diffusion coefficients $D_e $ and 
$D_i $ are calculated and shown in Fig.5.
Figure 5(a) corresponds to Fig.1(b).
Figure 5(b) corresponds to Fig.3(a).
The transport coefficients are estimated by taking the second order moment
of the test particles:
$D_{e,i} = (1/N) \sum_{i=1}^N [( \Delta r_{e,i} )^2 /2 \Delta t]$
where $\Delta r_{e,i}$ is  the radial excursion of the particles (the electrons or the ions)
measured with respect to the initial radial location. A sampling time is given by $ \Delta t$.
Here, $N$ is the number of test particles which are set on a same radial location
(and equidistantly in the poloidal angle).
Note that $D_e $ and $D_i$ represent transport at the transient phase and
not the ones at the steady state of a tokamak discharge.
Relatively short sampling times are set for the electrons [$\Delta t = 4.4 (\mu s) $]
and for the ions [$\Delta t = 0.27 (ms)$] because if the particles' spread
is too large, the information of the local stochasticity is lost.
Despite Fig.5 is generated by projecting the three dimensional
motion onto the radial coordinate, we can observe non-monotonic structure. 
The degree of particle stochasticity varies in space.
We again observe stepwise structure in the vicinity of $0.8 \le r/a \le 0.85$
for the ions (the $D_e$ behavior within $0.82 \le r/a \le 0.85$
remains yet to be explained).
The transport coefficients decrease near $r/a = 1.0$ 
due to the lower temperature of the test particles toward the edge.
As a reminder, the temperature profile is given by $T (r) = T_0 [1 - (r/a)^4]$ as an initial condition.

\begin{figure}
\includegraphics[width=72mm]{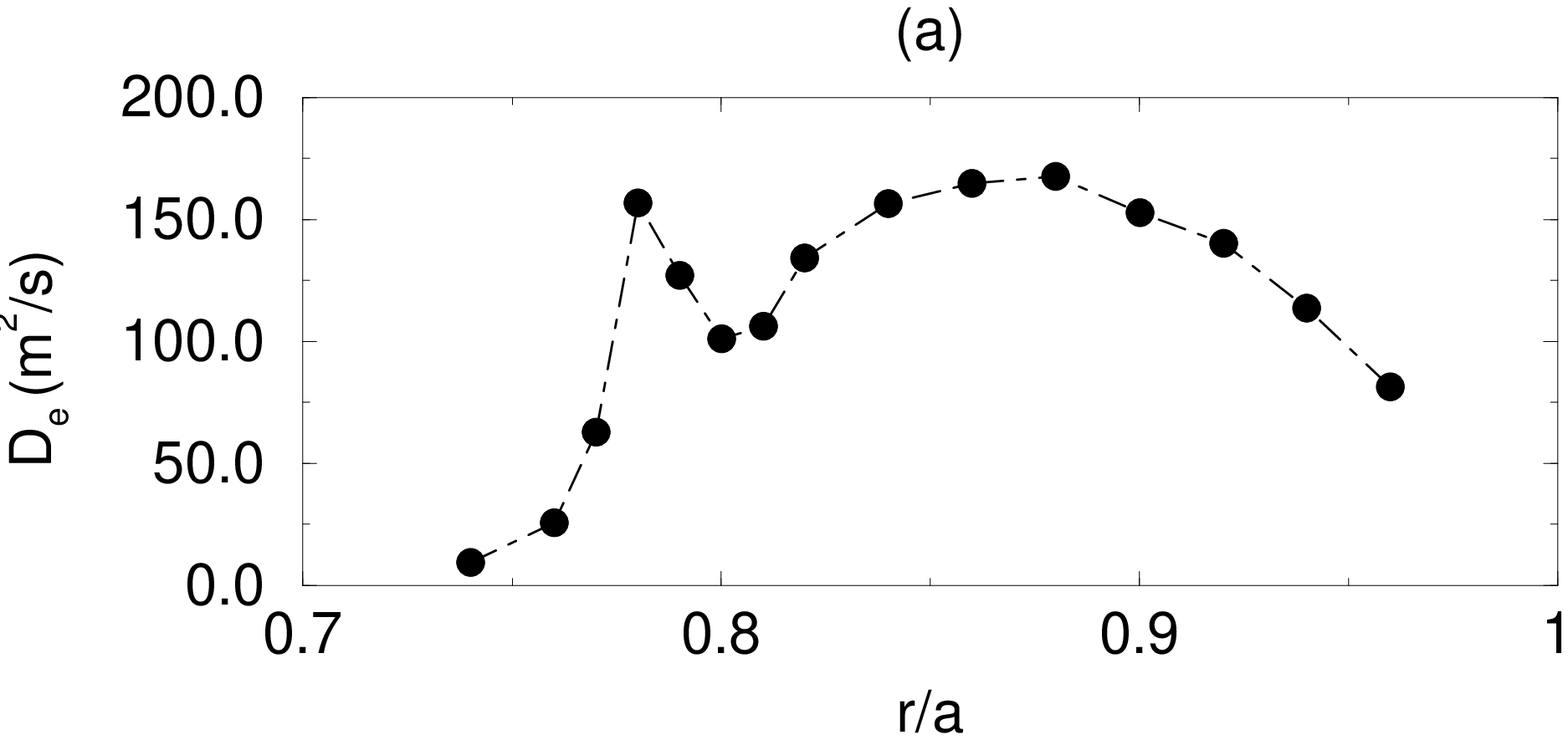}
\hfil
\includegraphics[width=72mm]{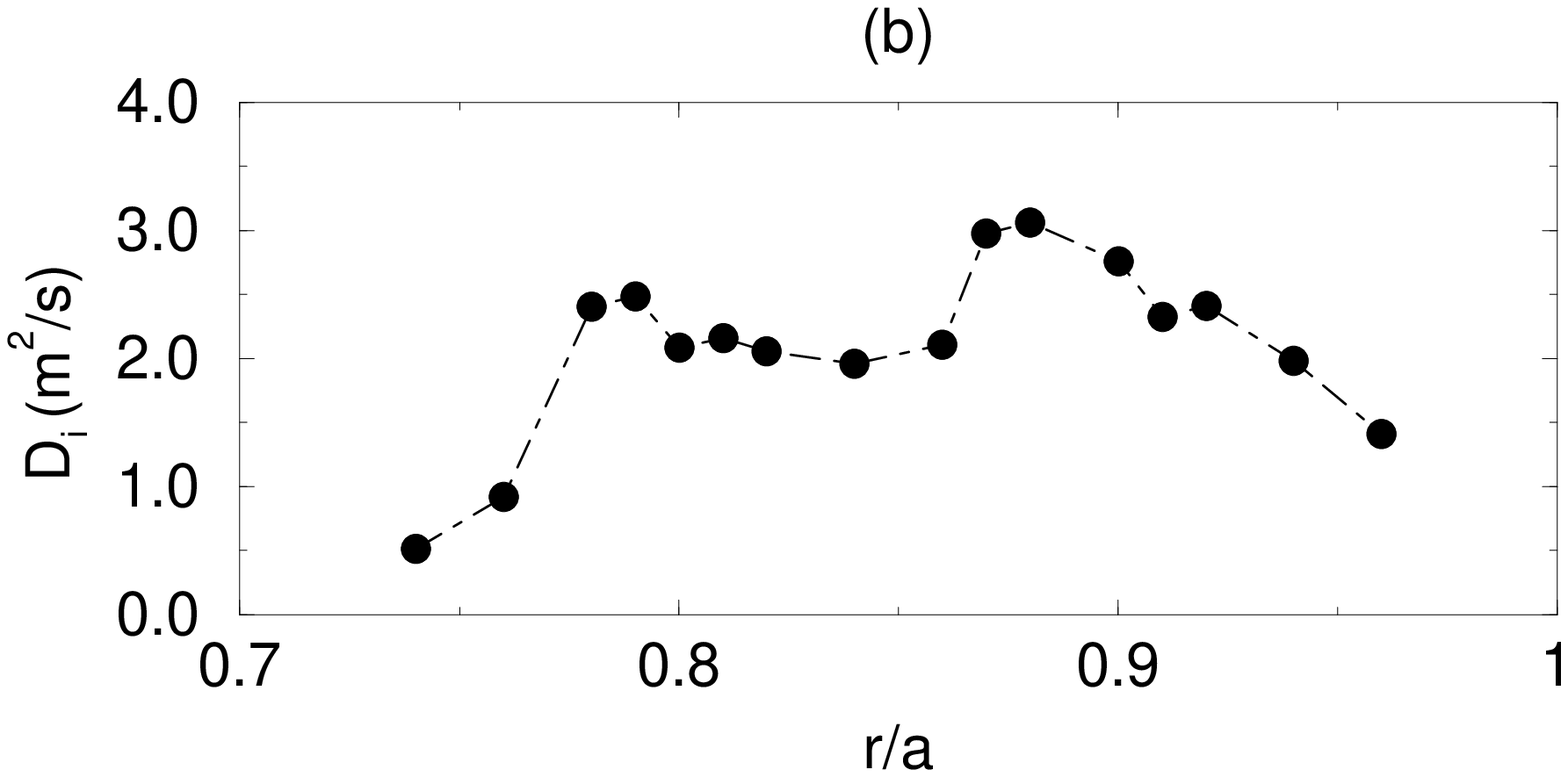}
\caption{(a) Electron diffusion coefficient $D_e   $ and 
(b) ion diffusion coefficient $D_i  $ as
a function of radial location. }
\label{fig5}
\end{figure}

\section{Summary and discussions}  
\label{s4}
In this work, we have investigated tokamak edge particle and heat transport
employing guiding center orbit following calculation in the presence of magnetic perturbation. 
From the orbit calculation, we have obtained the time evolution of radial density and temperature profile. 
Local diffusion coefficients are estimated by taking the second order moments of the test particles.

We have shown that the remnants of the magnetic islands can play
a role in characterizing the radial transport.
On the other hand, by raising the amplitude of the magnetic perturbation 
the stepwise structure disappeared accompanied by much rapid profile evolution.
From the profile evolution and the diffusion coefficients, the remaining magnetic island
regions can be regarded as a buffer for the radial transport
(provides particles with relatively good confinement region). 
The island remnants effect can be related
to the unresolved observation of the RMP experiments.\cite{eva06}
By increasing the trapped particle fraction, the 
transport level is reduced. The latter is related to the conservation of the second adiabatic invariants.\cite{lic92}
The projection of the particle dynamics onto the radial coordinate
resembles diffusion process.\cite{cha90}
In actuality, what we see is the radial projection of the freely streaming
electrons and ions along radially random walking stochastic magnetic field lines
(with the correction of the mirror force\cite{cha98} and the perpendicular drifts).
The latter effects appear as parallel convection in the fluid model
(the convective derivative originates from the streaming term in the Vlasov equation).
We further plan to analyze the transport process by the fluid model.\cite{cgl}

In this work, computational settings are not exactly that of
tokamak RMP experiments.\cite{eva06} 
For example, we did not have the divertor geometry. 
The safety factor, the density profile, and the temperature profile 
we used are given by a simplified model.
Orbit calculation employing more realistic parameters (of the DIII-D tokamak, for example)
will be our future work.
One of the authors (YN) would like to thank discussions
with Dr.~R.~A.~Moyer, Dr.~T.~E.~Evans, and Dr.~L.~E.~Zakharov.
This work is supported by National Science Council of Taiwan, NSC 100-2112-M-006-021-MY3.
Computation of this work is performed on NVIDA GPU cluster supported by  
National Cheng Kung University Top University Project and National Center for High-Performance
Computing (NCHC) ALPS cluster of Taiwan.



\begin{thebibliography}{[1]}

\bibitem{eva06}%
T.~E.~Evans, R.~A.~Moyer, K.~H.~Burrell, {\it et al.}, Nature Physics {\bf 2}, 419 (2006).

\bibitem{moy13}%
R.~A.~Moyer (private communication, 2013).

\bibitem{nf11}%
T.~W.~Petrie, T.~E.~Evans, N.~H.~Brooks, {\it et al.}, Nucl. Fusion {\bf 51}, 073003 (2011).

\bibitem{nf12}%
O.~Schmitz, T.~E.~Evans, M.~E.~Fenstermacher, {\it et al.}, Nucl. Fusion {\bf 52}, 043005 (2012).

\bibitem{rec78}%
A.~B.~Rechester and M.~N.~Rosenbluth, Phys. Rev. Lett. {\bf 40}, 38 (1978).

\bibitem{wes97}%
J.~A.~Wesson, B.~Alper, A.~W.~Edwards, and R.~D.~Gill, Phys. Rev. Lett. {\bf 79}, 5018 (1997); 
A rapid convective inward transport of nickel is reported. 

\bibitem{fur63}%
H.~P.~Furth, J.~Killeen, and  M.~N.~Rosenbluth, Phys. Fluids {\bf 6}, 459 (1963).

\bibitem{miy86}%
K.~Miyamoto, {\it Plasma Physics for Nuclear Fusion}, 2nd ed. (MIT Press, Cambridge, 1989), P.41.  

\bibitem{nis11}%
Y.~Nishimura, Comput. Phys. Commun. {\bf 182}, 158 (2011);
we initially focus on the collision-less particles.

\bibitem{cha90}%
Z.~Chang and J.~D.~Callen, Nucl. Fusion {\bf 30}, 219 (1990).

\bibitem{nic92}%
D.~R.~Nicholson, {\it Introduction to Plasma Theory}, 2nd ed. (Kreiger, Malabar, 1992), P.92.

\bibitem{cha98}%
B.~Chandran and S.~Cowley, Phys. Rev. Lett. {\bf 80}, 3077 (1998).

\bibitem{lic92}%
A.~J.~Lichtenberg and M.~A.~Lieberman, {\it Regular and Chaotic Dynamics}, 2nd ed. (Springer, Berlin, 1992), P.21.  

\bibitem{cgl}%
G.~F.~Chew, M.~L.~Goldberger, and F.~E.~Low, Proc. Royal Soc. London {\bf A 236}, 112 (1956). 

\end{thebibliography}
\end{document}